\documentstyle[prl,aps,floats,psfig]{revtex}   
\begin{document}
\twocolumn[
\hsize\textwidth\columnwidth\hsize\csname@twocolumnfalse\endcsname

\title{Proposal for all-electrical measurement of T$_1$ in semiconductors} 

\author{Igor \v{Z}uti\'{c}$^1$, Jaroslav Fabian$^2$, and 
S. Das Sarma$^{1}$} 
\address{ $^1$Condensed Matter Theory Center, Department of Physics, 
University of Maryland, 
College Park, Maryland 20742-4111 \\
$^2$Institute for Theoretical Physics, Karl-Franzens 
University, Universit\"{a}tsplatz 5, 8010 Graz, Austria}

\maketitle
 
\begin{abstract}
In an inhomogeneously doped magnetic semiconductor spin
relaxation time $T_1$ can be determined by {\it all-electrical}
measurements. Nonequilibrium spin injected in a magnetic {\it p-n}
junction gives rise to the spin-voltaic
effect where the nonequilibrium spin-induced charge current is very sensitive
to $T_1$ and can flow even at no applied bias. It is proposed that
$T_1$ can be determined by measuring the I-V characteristics in
such a geometry. For a magnetic {\it p-n} junction where  the results
can be calculated analytically, it is in addition  possible to extract the
g-factor and the degree of injected carrier spin polarization. 
\end{abstract}
\vspace{-0.6cm}
]

In examining the properties of spin-polarized transport in solid state
systems one of the key  physical quantities is  the characteristic 
spin relaxation time $T_1$ and the related length scale, spin 
diffusion length $L_s$, both describing the decay of  nonequilibrium spin.
These spin relaxation parameters play crucial roles in various 
novel 
spintronic applications~\cite{dassarma01}. Unlike in the conventional 
charge-based electronics, spintronic devices rely on  manipulating  
nonequilibrium spin. 
Since $T_1$ and $L_s$  determine ``spin memory'' they
effectively set an upper limit on the time required to perform various 
device operations and the possible optimal size of  spintronic devices.
In semiconductor
spintroncs~\cite{dassarma01}, 
spin relaxation of carriers (electrons and holes) is a complex 
process~\cite{optical84,fabian99}. 
For a given temperature and doping, several different
mechanisms  contribute to spin relaxation which is 
sensitive~\cite{optical84,fabian99} 
to strain, dimensionality,  magnetic and electric fields. 
It  would be highly desirable if the same semiconductor structures which
hold promise for spintronic applications could also be used to probe 
spin relaxation. Previous methods~\cite{optical84,fabian99} to measure $T_1$
have typically used optical techniques or electron spin resonance.

In this letter we discuss a proposal to determine $T_1$ by {\it all-electrical}
measurements from the I-V characteristics. This method can be viewed as a 
generalization of the concept of {\it spin-charge} 
coupling~\cite{silsbee80,johnson85}, 
introduced in metals by Silsbee and Johnson, 
to inhomogeneously doped semiconductors~\cite{zutic02}.
We show how several
features, specific to semiconductors (bipolar transport--by both electrons
and holes, bias-dependent depletion region, and highly nonlinear I-V
characteristics), can be exploited to provide a sensitive probe for $T_1$. 

\begin{figure}
\centerline{\psfig{file=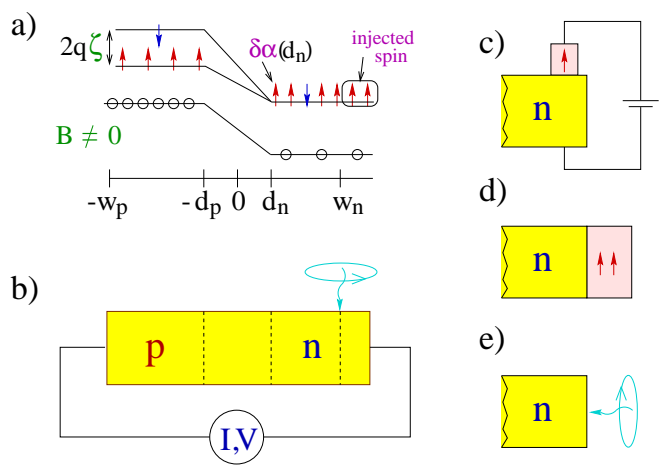,width=1.0\linewidth,angle=0}} 
\vspace{0.6truecm}
\caption{Scheme of a magnetic {\it p-n} junction. a) Band-energy diagram with 
spin-polarized electrons (arrows) and unpolarized holes (circles).
b) Circuit geometry corresponding to panel a). Using circularly
polarized light, nonequilibrium spin is injected transversely in the 
nonmagnetic $n$ region and the circuit loop for  I-V characteristics is indicated. 
Panels c) - e) indicate alternative schemes to inject spin into the  $n$ region.
Schemes c) and d) rely on the magnetic (paramagnetic or ferromagnetic) material
to inject spin electrically. Realizations depicted in  b), c), and e) are suitable to 
demonstrate spin-voltaic effect,
where: 
1) in a closed circuit charge current can flow, even at no applied (longitudinal) bias,
with the direction which can be reversed either by $B\rightarrow -B$ or by 
the reversal of the orientation of the injected spin,
2) for an open circuit an analogous reversal in $B$ or in the spin orientation 
would change the sign of the voltage drop across the junction. }
\vspace{-0.8truecm}
\label{fig:1}
\end{figure}

To illustrate our proposal we consider a magnetic {\it p-n} junction~\cite{zutic02,fabian02} 
as sketched in Fig.~\ref{fig:1}a,b. 
In the $p$ ($n$) region there is a 
uniform doping with $N_a$ acceptors ($N_d$ donors). Within the  depletion
region ($-d_p < x < d_n$) we assume that there is a spatially dependent 
spin splitting of the carrier bands.
Such  splitting, a consequence of doping with magnetic impurities,  can 
occur in different situations. For example, in ferromagnetic 
semiconductors~\cite{ohno98} or, 
in the presence of magnetic field $B$, 
the spin splitting could arise from 
 either having inhomogeneous $g$-factors 
or by applying an inhomogeneous magnetic field. While our method is applicable to 
all of these cases, we focus here on the last two instances and further assume that
the carriers obey the nondegenerate Boltzmann statistics. 
In the low injection regime it is possible to obtain the results  for 
spin-polarized transport analytically and to decouple the contribution of electrons 
and holes~\cite{fabian02}. Following the approach from 
Ref.~\cite{zutic02}, we consider only the effect of spin-polarized electrons. 
(It is simple to  also include the net spin polarization of holes~\cite{fabian02}).   
The resulting 
Zeeman splitting of the conduction  band (Fig.~\ref{fig:1}a)  is 
$2  q\zeta=g\mu_B B$,
where $g$ is the $g$-factor for electrons, 
$\mu_B$ is Bohr magneton, $q$ is the proton charge, and
$\zeta$ is the electron  magnetic potential~\cite{zutic02}. 

Nonequilibrium electron and hole densities are $n$ (the sum of spin up and
spin down components $n_{\uparrow}+n_{\downarrow}$)
and $p$, while the 
spin density and its  polarization are  $s=n_\uparrow-n_\downarrow$ and 
$\alpha=s/n$, respectively.
Equilibrium values (with subscript ``0'') satisfy
$n_0 p_0=n_i^2\cosh(\zeta/V_T)$ and $\alpha_0=\tanh(\zeta/V_T)$, 
where $n_i$ is the
intrinsic (nonmagnetic) carrier density and $V_T=k_B T/q$, with $k_B$ being 
the Boltzmann constant and $T$ temperature. 
We assume~\cite{zutic02} equilibrium values (ohmic contacts) 
for minority carriers  at $x=-w_p,w_n$ and at $x=-w_p$ for spin density. 
To characterize the spin injection, at $x=w_n$ we impose $\delta s(w_n)=\alpha(w_n) N_d$, 
where $\delta s=s-s_0$  and $\delta \alpha=\alpha-\alpha_0$.  
(Neglecting $\delta p(w_n)$, which can accompany $\delta s(w_n)$, 
is an accurate approximation while ($w_n - d_n$) is greater than the hole diffusion length~\cite{fabian02}.) 
In addition to spin injection 
by optical means~\cite{optical84}
(depicted in Fig.~\ref{fig:1}b,e), an electrical spin injection
(Fig.~\ref{fig:1}c,d) has been reported  using a wide  range of magnetic 
materials~\cite{fiederling99,ohno99,jonker00,hammar01,isakovic01,jonker01}.
For a magnetic {\it p-n} junction total charge current $J$ can be 
decomposed~\cite{zutic02,fabian02} as the sum of 
equilibrium-spin electron $J_n$ and
hole $J_p$ currents, and spin-voltaic current $J_{sv}$,
which originates from the interplay of the equilibrium magnetization
(i.e. equilibrium spin polarization in the $p$ region) and the 
nonequilibrium spin (injected in the $n$ region).  

The individual contributions of $J$  
as a function of applied bias 
$V$ and $B$ (recall that  $\zeta=\zeta(B)$) are~\cite{zutic02,fabian02} 
\begin{eqnarray} \label{eq:J1}
J_{n}&=&q\frac{D_n}{L_n} n_0(-d_p)
 \coth\left (\frac{\tilde{w}_p}{L_n}\right) \left (e^{V/V_T}-1\right ),\\ \label{eq:J2}
J_{p}&=&q\frac{D_p}{L_p} p_0(d_n) \coth\left (\frac{\tilde{w}_n}{L_p}\right ) 
\left (e^{V/V_T}-1\right ),\\ \label{eq:J3}
J_{sv}&=&q \frac{D_n}{L_n} n_0(-d_p) \coth\left (\frac{\tilde{w}_p}{L_n}\right ) 
e^{V/V_T} \alpha_0(-d_p) \delta \alpha(d_n),
\end{eqnarray}
where $D_n$ ($D_p$) is the electron (hole) diffusivity, 
$L_n$ and $L_p$
are the minority diffusion lengths~\cite{zutic01}, 
and $\tilde{w}_p=w_p-d_p$ ($\tilde{w}_n=w_n-d_n$) is the width 
of the bulk $p$ ($n$) region. There is an implicit $V$-dependence
of  $\tilde{w}_{n,p}$ since for the depletion layer edge~\cite{ashcroft76} 
$d_{n,p}\propto \sqrt{V_b-V}$, where $V_b=V_T \ln(N_a N_d/n_i^2)$ is
the built-in voltage. The derivation of the Eqs.~\ref{eq:J1}-\ref{eq:J3}
assumes that the depletion region is highly resistive (depleted from
free carriers)~\cite{fabian02,ashcroft76}.  
The voltage drop 
between the two ends of the junction (see Fig.~\ref{fig:1}) and between 
$x=-w_p$ and $x=w_n$ can then be identified. 

\begin{figure}
\centerline{\psfig{file=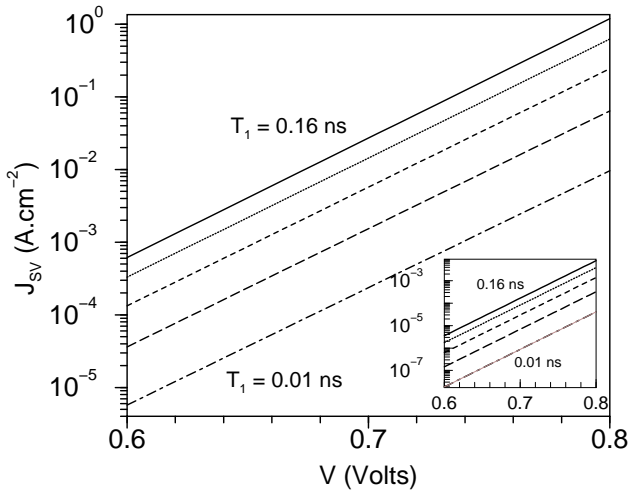,width=1.0\linewidth,angle=-90}} 
\vspace{0.5truecm}
\caption{Calculated  spin-voltaic current $J_{sv}$ for the magnetic {\it p-n}
junction as a function of forward bias (in Volts).
Lines (top to bottom) correspond to $T_1=0.16,0.08,0.04,0.02,$ and $0.01$ ns,
revealing the high sensitivity for probing the spin relaxation time. 
The doping is $N_a=N_d=5\times 10^{15}$ cm$^{-3}$. In the inset the 
results are displayed for $N_a=N_d=5\times 10^{17}$ cm$^{-3}$
(all the other parameters remain unchanged), indicating that the 
high sensitivity to $T_1$ is preserved at different doping levels.
}
\label{fig:2}
\vspace{-0.3cm}
\end{figure}

We next explore some properties of charge current which will be used 
to formulate the method for determining $T_1$.  
From Eq.~\ref{eq:J3} we note $J_{sv} \propto \delta \alpha(d_n)$, the 
spin-voltaic part of the charge current is related to the nonequilibrium spin.
For a given injected spin, represented by $\delta \alpha(w_n)$, it follows
(see Fig.~\ref{fig:1}a) that $J_{sv}$ should be sensitive to:
1) $\tilde{w}_n$ the separation between the source of spin injection and the 
depletion layer edge, and 2) the  spin diffusion length $L_{sn}=\sqrt{D_nT_1}$, 
characterizing the spin decay, i.e., $\delta \alpha(w_n)$. 
Indeed, one can show~\cite{zutic02} that 
\begin{equation} 
\delta \alpha(d_n)=\delta \alpha(w_n) / \cosh (\tilde{w}_n/L_{sn}), 
\label{eq:alpha}
\end{equation}
which from Eq.~\ref{eq:J3} implies a high sensitivity of $J_{sv}$ to $T_1$
(through $L_{sn}$).
In contrast, $J_{n,p}$ do not contain the nonequilibrium spin  and thus
have no $T_1$ dependence.
A direct measurement of total charge current to identify $T_1$ 
[based on $J_{sv}=J_{sv}(T_1)$] implies some limitations. At vanishing
bias ($V \ll V_T$), where $J_{n,p}\rightarrow 0$, $J\rightarrow J_{sv}$ is
small, while at higher bias ($V \gg V_T$ and $V < V_b$) $J$ is dominated
by $J_n$ and $J_p$--large $T_1$-independent background.
To fully exploit simple I-V measurements we note that $T_1=T_1(|B|)$
(the precise $B$-dependence differs for various spin-relaxation mechanisms).
We also use the symmetry properties of the individual contributions to the 
charge current
with respect to the applied magnetic field: $J_{n,p}(-B)=J_{n,p}(B)$, and
$J_{sv}(-B)=-J_{sv}(B)$. This follows if we recall that $\zeta \propto B$,
$J_n \propto \cosh(\zeta/V_T)$, $J_p$ is $\zeta$-independent, and 
$J_{sv}\propto \sinh(\zeta/V_T)$. Consequently, 
by measuring $J(V,B)-J(V,-B)=2J_{sv}$
the large $T_1$-independent background has then been effectively removed. 
To optimize the experimental sensitivity we assume that, with the exception of $T_1$,
all the  material parameters are known and consider variable sample size which 
would give large difference in $J_{sv}$ as $T_1$ is changed, i.e., large 
$\partial[\delta \alpha(d_n)]/\partial L_{sn}$ (see Eq.~\ref{eq:alpha}). 
For a given $L_{sn}$ this is achieved with $\tilde{w}_n/L_{sn} \approx 1.5$ 
and to increase the magnitude of $J_{sv}$ it is favorable to choose a short 
$p$ region~\cite{short} 
($J_{sv}\propto \coth (\tilde{w}_p/L_n)$) and 
to consider forward bias
$V \gg V_T$, while still remaining in the low bias (low injection) regime 
($V<V_b$). 
Since a priori we can only estimate a range of expected values for $T_1$ 
the choice of $\tilde{w}_n$ should maximize the corresponding values of 
$\partial[\delta \alpha(d_n)]/\partial L_{sn}$. 
The results obtained by this procedure are illustrated in Fig.~\ref{fig:2}. 

The material parameters are based on 
GaAs~\cite{zutic01} 
$D_n=10D_p=103.6$ cm$^2$s$^{-1}$, $L_n\approx 1.0$ $\mu$m, 
$L_p\approx 0.3$ $\mu$m,
$n_i=1.8 \times 10^6$ cm$^{-3}$. Doping with 
$N_a=N_d=5\times 10^{15}$ cm$^{-3}$ at $V=0$ yields $d_n=d_p\approx 0.4$ $\mu$m.  
For example, expecting that the spin relaxation time will be within  
$0.01$ and $0.16$ ns, to optimize sensitivity,  we choose that 
for $T_1=0.16$ ns (which corresponds to $L_{sn}\approx 1.3$ $\mu$m)  
$\tilde{w}_n/L_{sn}\approx 1.5$. We set (at $V=0$) 
$\tilde{w}_p\approx 0.3$ $\mu$m, which leads (see Fig.~\ref{fig:1}a,b)
to $w_p=0.7$ $\mu$m, $w_n=2.3$ $\mu$m. 
We use for the injected spin polarization $\delta \alpha(w_n)=0.5$
and for the maximum spin splitting $2 q \zeta/V_T=0.2$, where
at room temperature and B [Tesla] one can also write  
$q \zeta/V_T \approx 900 B/g$~\cite{zutic02}.
The sensitivity of our methods is displayed in Fig.~\ref{fig:2}
where for approximately an order of magnitude change in $T_1$ the spin-voltaic
current $J_{sv}$ changes by {\it two} orders of magnitude.
Since the gist of the method outlined above relies on the robust symmetry 
properties of
$J_{n,p}$ and $J_{sv}$ with respect to $B$, it is straightforward to implement
our proposal for a wide variety of magnetic {\it p-n} junctions where only a 
numerical solution is known. 
For example, higher (degenerate) doping could also be considered, typical
for ferromagnetic semiconductors~\cite{ohno98}.

With the aid of the analytic solution of Eqs.~\ref{eq:J1}-\ref{eq:J3} it is also
possible to illustrate how to extract other quantities of interest. Consider 
the situation where we accurately know $B$ and are interested in 
measuring $g$-factor 
in the magnetic $p$ region. Recalling that $2 q\zeta=g\mu_B B$, identifying 
$\zeta$ is then equivalent to extracting the $g$-factor. We use that
$J_{n,p}$ is even in $B$ and measure $J(V,B)+J(V,-B)=2[J_n(V,B)+J_p(V,B)]$.
From  Eqs.~\ref{eq:J1} and \ref{eq:J2} we note that 
the only dependence on $\zeta$ ($\propto B$) enters through
$n_0(-d_p)=(n_i^2/N_a)\cosh(\zeta/V_T)$. 
Consequently, $J(V,B)+J(V,-B)\equiv a(V) + b(V)\cosh(\zeta/V_T)$, 
where functions a(V), b(V) are known and readily expressed in terms of the parameters 
from Eqs.~\ref{eq:J1} and \ref{eq:J2}.
It remains then to measure $J(V,B)+J(V,-B)$ for different values of $B$ and to obtain a one parameter 
fit for $\zeta$, i.e., for the $g$-factor. 
(An attempt to extract $\zeta$ from $J_{sv}$ would be more complicated,
since it also contains, generally unknown, $B$-dependence in $T_1$.)
If both $\zeta$ and $T_1$ are unknown, this procedure to obtain $\zeta$ 
should then be followed by measuring the spin-voltaic current to extract $T_1$, 
as discussed above. 
Finally, we  consider the situation where $\zeta$, $T_1$, and 
$\delta \alpha(w_n)$  
are all unknown. 
Again we first extract $\zeta$ from  $J(V,B)+J(V,-B)$
and subsequently the spin-voltaic current by measuring $J(V,B)-J(V,-B)$.
It follows from Eq.~\ref{eq:J3} that the value of $\delta \alpha(d_n)$ 
can then be determined.
However,  $\delta \alpha(d_n)$ 
(see Eq.~\ref{eq:alpha}) still contains two unknown quantities: 
$T_1$ ($L_{sn}$) and $\delta \alpha(w_n)$ which influence
needs to be decoupled. We assume  (as it was implicitly done throughout the paper) 
that $\delta \alpha(w_n)$ is $V$-independent. We recall that change of
applied bias modifies  $d_n$. 
Effectively, we are changing the separation between the point of spin injection and 
spin ``detection,'' since at the depletion edge $x=d_n$ the remaining nonequilibrium spin 
can be detected by its measurable effect on charge current. To eliminate influence of
$\delta \alpha(w_n)$ we  evaluate 
$f(V)=\delta \alpha[d_n(V_0)]/\delta \alpha[d_n(V)]$ for a range of applied bias
and fixed $V_0$. In our case, it is suitable to choose $V \in [-0.8,0.8]$ and $V_0=-0.8$.
Variation of $f(V)$ changes monotonically with $T_1$ and can be used to extract 
the spin relaxation time. However, the resulting sensitivity will be smaller that
the one achieved in Fig.~\ref{fig:2}, under the assumption that $T_1$ is the
only unknown quantity. 
[For $T_1=0.16$ ns 
$\Delta f(V)\equiv [f(0.8)-f(-0.8)]/f(-0.8) \approx 0.25$, while for $T_1=0.01$ ns
$\Delta f(V) \approx 1.7$.] After $T_1$ is extracted we use then Eq.~\ref{eq:alpha}
to obtain $\delta \alpha(w_n)$,
the only remaining unknown quantity.

We have proposed here how {\it all-electrical} measurements can be used to 
identify several quantities fundamental to the understanding of spin-polarized
transport in semiconductors. The general principle that the nonequilibrium 
injected spin can produce measurable effects on charge current should be
useful both for developing novel device concepts in semiconductor spintronics,
as well as a diagnostic tool for the existing structures. 

This work was supported by DARPA, NSF, and the US ONR.


\vspace{-0.4cm}
\begin{references}
\vspace{-1.5cm}

\bibitem{dassarma01} 
S. Das Sarma, J. Fabian, X. Hu, and 
I. \v{Z}uti\'{c}, IEEE Transaction on Magnetics {\bf 36}, 
2821 (2000); Superlattice Microst. 
{\bf 27}, 289 (2000);  Solid State Commun. {\bf 119}, 207 (2001).

\bibitem{optical84}
F. Meier and B. P. Zakharchenya (eds),
{\it Optical Orientation} (North-Holland, New York 1984).

\bibitem{fabian99}
J. Fabian and S. Das Sarma, J. Vac. Sci. Technol. B {\bf 17}, 1780 (1999).                                                                
\bibitem{silsbee80}
R. H. Silsbee, Bull. Magn. Reson. {\bf 2}, 284 (1980).
 
\bibitem{johnson85}
M. Johnson and R. H. Silsbee, Phys.
Rev. Lett. {\bf 55}, 1790 (1985).                                  

\bibitem{zutic02} 
I. \v{Z}uti\'{c}, J. Fabian, and S. Das Sarma,
Phys. Rev. Lett. {\bf 88}, 066603 (2002).

\bibitem{fabian02} 
J. Fabian, I. \v{Z}uti\'{c}, and S. Das Sarma, preprint.

\bibitem{ohno98}
H. Ohno, Science {\bf 281}, 951 (1998).

\bibitem{ashcroft76} N. W. Ashcroft and N. D.  Mermin, {\it Solid
State Physics} (Saunders, New York, 1976).
   
\bibitem{fiederling99} 
R. Fiederling, M. Kleim, G. Reuscher, W. Ossau, G. Schmidt, A. Waag, 
and L. W. Molenkamp,  Nature {\bf 402}, 787 (1999).

\bibitem{ohno99} 
Y. Ohno, D. K. Young, B. Beschoten, F. Matsukura, H. Ohno, and
D. D. Awschalom, Nature {\bf 402}, 790 (1999).

\bibitem{jonker00} 
B. T. Jonker, Y. D. Park, B. R. Bennett, H. D. Cheong,
G. Kioseoglou, and A. Petrou, Phys. Rev. B {\bf 62}, 8180 
(2000).

\bibitem{hammar01}
P. R. Hammar and M. Johnson,
Appl. Phys. Lett. {\bf 79}, 2591 (2001).
                                                                              
\bibitem{isakovic01}
A. F. Isakovi\'{c}, D. M. Carr, J. Strand, B. D. Schultz, C. J. Palmstr{\o}m,
and P. A. Crowell, Phys. Rev. B {\bf 64}, 161304 (2001).                                         

\bibitem{jonker01}
A. T. Hanbicki, B. T. Jonker, G. Itskos, G. Kioseoglou, and A. Petrou,
Appl. Phys. Lett. {\bf 80}, 1240 (2002). 

\bibitem{zutic01} 
I. \v{Z}uti\'{c}, J. Fabian, and S. Das Sarma, Phys. Rev. B {\bf 64}, 121201 (2001); 
Appl. Phys. Lett. {\bf 79}, 1558 (2001).

\bibitem{short} $\tilde{w}_p > 0$,
for ohmic boundary conditions to be accurate. 
\end{references}
\end{document}